\documentclass[aps,prb,showpacs,twocolumn,superscriptaddress,floatfix]{revtex4}
\usepackage{epsfig,amsmath,amssymb}
\def\beq{\begin{equation}}
\def\eeq{\end{equation}}
\def\bea{\begin{eqnarray}}
\def\eea{\end{eqnarray}}
\def\nn{\nonumber}
\def\abs#1{\left\vert#1\right\vert}

\begin{document}

\title{Quantum kagom\'e antiferromagnet in a magnetic field:
Low-lying non-magnetic excitations versus valence-bond crystal order}
\author{D.C.\ Cabra}
\affiliation{Universit\'{e} Louis Pasteur,
  Laboratoire de Physique Th\'{e}orique,
  67084 Strasbourg C\'edex, France}
\author{M.D.\ Grynberg}
\affiliation{Departamento de F\'{\i}sica, Universidad Nacional de
La Plata,  C.C.\ 67, (1900) La Plata, Argentina}
\author{P.C.W.\ Holdsworth}
\affiliation{Laboratoire de Physique,
      ENS Lyon, 46 All\'ee d'Italie, 69364 Lyon C\'edex 07, France}
\author{A.\ Honecker}
\affiliation{TU Braunschweig, Institut f\"ur Theoretische Physik,
  38106 Braunschweig, Germany}
\affiliation{Universit\"at Hannover, Institut f\"ur Theoretische
  Physik, Appelstrasse 2, 30167 Hannover, Germany}
\author{P.\ Pujol}
\affiliation{Laboratoire de Physique,
      ENS Lyon, 46 All\'ee d'Italie, 69364 Lyon C\'edex 07, France}
\author{J.\ Richter}
\affiliation{Institut f\"ur Theoretische Physik, Otto-von-Guericke
 Universit\"at Magdeburg,
 39016 Magdeburg, Germany}
\author{D.\ Schmalfu{\ss}}
\affiliation{Institut f\"ur Theoretische Physik, Otto-von-Guericke
 Universit\"at Magdeburg,
 39016 Magdeburg, Germany}
\author{J.\ Schulenburg}
\affiliation{Universit\"atsrechenzentrum, Otto-von-Guericke-Universit\"at
 Magdeburg,
 39016 Magdeburg, Germany}

\date{April 13, 2004; revised December 22, 2004}

\begin{abstract}
We study the ground state properties of a quantum antiferromagnet
on the kagom\'e lattice in the presence of a magnetic field,
paying particular attention to the stability of the plateau at
magnetization $1/3$ of saturation and the nature of its ground
state. We discuss fluctuations around classical ground states
and argue that quantum and classical calculations at the harmonic level do not
lead to the same result in contrast to the zero-field case. For spin $S=1/2$
we find a magnetic gap below which an exponential number of non-magnetic
excitations are present. Moreover, such non-magnetic excitations
also have a (much smaller) gap above the three-fold degenerate ground state.
We provide evidence that the ground state has long-range order of
valence-bond crystal type with nine spins in the unit cell.
\end{abstract}


\maketitle


\section{Introduction}

The appearance of exotic quantum phases in systems described by
two-dimensional frustrated antiferromagnets is presently the subject
of intense research (see e.g.\ Refs.\ \onlinecite{lhuillier03,RSH04} for
recent reviews). The Heisenberg antiferromagnet on highly frustrated
lattices such as the pyrochlore and kagom\'e lattice has a huge
degeneracy of the classical ground state such that no magnetic
order arises at any temperature (see e.g.\ Ref.\ \onlinecite{mRev01} for
a recent review). At the quantum level one may then obtain different
exotic phases without magnetic (N\'eel) order. One such phase is
the so-called `valence-bond crystal' which is characterized by
formation of local singlets in a long-range ordered pattern.
An even more exotic phase, namely one without any kind of long-range
order, is suspected to arise in the $S=1/2$ Heisenberg model on
the kagom\'e lattice.%
\cite{lecheminant97,waldtmann98,Mila,lhuillier00,lhuillier03}
In the latter case, there is a small spin gap and,
although this is still under discussion,\cite{syroma02,NiSe03}
the ground state is suspected to be disordered. In particular, a huge
number of singlets (exponentially growing with the system size)
is found inside the spin gap which are reminiscent of the classical
degeneracy.

The spin $S=1/2$ kagom\'e Heisenberg antiferromagnet (KHAFM) is realized
e.g.\ in volborthite,\cite{kago} although presumably in some distorted
form. Another possible realization is given by atomic quantum gases
in optical lattices.\cite{qopt} In the latter case, magnetization
corresponds to particle number and a magnetic field to chemical
potential, opening the possibility to perform experiments for the
behavior of the spin model in a magnetic field.

\begin{figure}
\centerline{\psfig{file=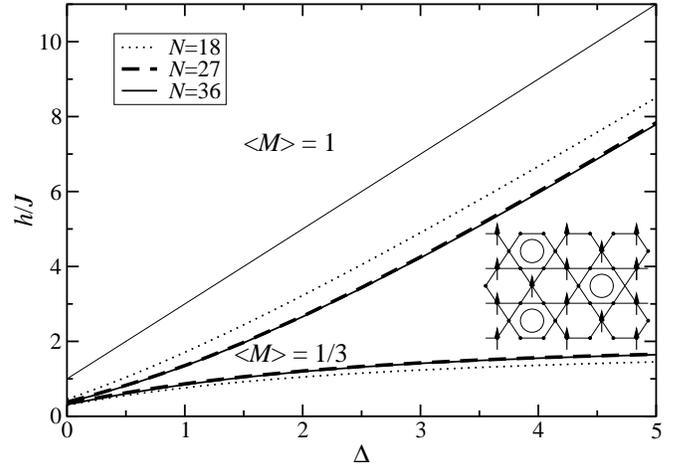,width=\columnwidth}}
\caption{Boundaries of the $\langle M\rangle=1/3$ plateau as a function of the
anisotropy $\Delta$ for different lattice sizes
(see legend) and the transition
to saturation $\langle M\rangle=1$ for the thermodynamic limit
(thin full line).
Inset: kagom\'e lattice with an ordered state of the valence-bond
crystal type at $\langle M \rangle = 1/3$:
circles in certain hexagons indicate local resonances between
different N\'eel configurations on the hexagons, arrows indicate
spins which are aligned with the field.
\label{figKagM1o3} }
\end{figure}

The magnetization process of the KHAFM 
has been studied theoretically both for classical \cite{Zhito,HPZ01,cghp}
as well as quantum spins.\cite{hida,cghp,SHSRS02,HSR03}
Numerical results for the magnetization curve of the $S=1/2$
Heisenberg model exhibit among others a clear plateau at $1/3$ of the
saturation magnetization \cite{hida,cghp,SHSRS02,HSR03} (see also Fig.\
\ref{figKagM1o3}). For the classical KHAFM 
at one third of the saturation field
thermal fluctuations select collinear states, but
there appears to be no real order.\cite{Zhito} For the $S=1/2$
KHAFM 
we will argue in this paper that the state with
magnetization $\langle M \rangle=1/3$ exhibits order of the
valence-bond crystal type (the spin-spin correlation functions are
short-ranged\cite{SCESproc}) although it shares some similarities with the
case $\langle M \rangle=0$.

In the present paper we study the $XXZ$ model in a magnetic field $h$
\begin{equation}
H = J \sum_{\langle i, j \rangle} \left(s^x_i s^x_j + s^y_i s^y_j
+ \Delta s^z_i s^z_j \right) - h \sum_i s^z_i \, ,
\label{HXXZ}
\end{equation}
where $\langle i, j \rangle$ indicates nearest neighbors on the
kagom\'e lattice (see inset of Fig.\ \ref{figKagM1o3}), $s_i^\alpha$
are spin-half operators acting at site $i$ and $\Delta$ is the
$XXZ$-anisotropy.

\begin{figure}
\centerline{\psfig{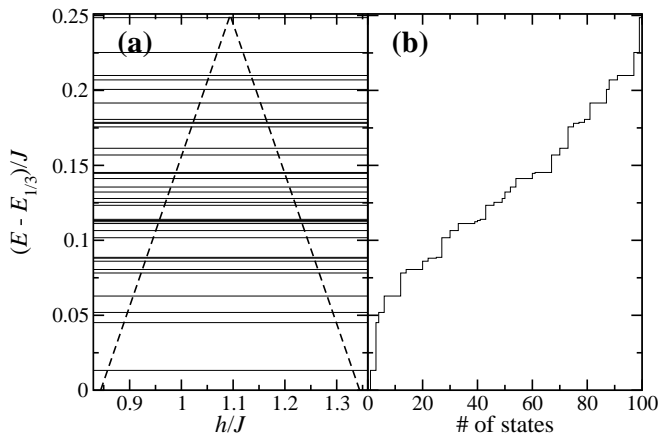}} \caption{
Low-lying excitations above the $\langle M \rangle = 1/3$ plateau
for the $S=1/2$ Heisenberg antiferromagnet
on the $N=36$ kagom\'e lattice. (a) Full lines show all excitations with
$S^z = 6$ in the given energy range, bold dashed lines the lowest
excitations with $S^z = 5$ and $S^z = 7$ as a function of magnetic
field $h$. (b) Excitation energy versus number of states with $S^z
= 6$ below that energy. One observes a total of 100 states below
the magnetic gap in the middle of the $\langle M \rangle = 1/3$
plateau (corresponding to the largest gap to magnetic
excitations).  \label{figLL36} }
\end{figure}

\section{Exact diagonalization for the $S=1/2$ Heisenberg model}

First we present numerical results which have
been obtained by Lanczos diagonalization of the Hamiltonian (\ref{HXXZ})
using the program package {\tt spinpack}.\cite{spinpack} All numerical
computations have been performed on lattices with $N$ sites subject to
periodic boundary conditions.

The main panel of Fig.\ \ref{figKagM1o3} shows the boundaries of the
fully polarized state (which we normalize to $\langle M \rangle = 1$)
and a state with $\langle M\rangle=1/3$ in the $XXZ$ model (\ref{HXXZ}).
Fig.\ \ref{figLL36}(a) shows the low-lying excitations above
the $\langle M \rangle =1/3$ ground state computed by exact diagonalization
for $N=36$ and $\Delta = 1$. The dashed lines show the gap to states with
$S^z = 5$ and $7$ which vanishes at the boundaries of the plateau.
The maximum magnetic gap occurs in the middle of the plateau where these
two lines intersect and we will use this as a definition of the magnetic gap.
Horizontal straight lines denote states with $S^z = 6$ ({\it i.e.}\
$\langle M \rangle =1/3$) and correspond to non-magnetic excitations.
The large number of non-magnetic excitations below the magnetic
gap is reminiscent of the classical degeneracy.
The shape of the integrated density of non-magnetic excitations (see
Fig.\ \ref{figLL36}(b)) is very similar to the corresponding
integrated density of singlets at $\langle M \rangle = 0$ (see
Fig.\ 2 of Ref.\ \onlinecite{waldtmann98}). In combination with the disordered
classical ground state,\cite{Zhito} one might be tempted to take this
as evidence that also the ground state of the $S=1/2$ KHAFM 
at $\langle M \rangle = 1/3$ is disordered.
However, we will argue next that here classical and quantum fluctuations
are in fact not equivalent at the harmonic level and then present
evidence in favor of an {\it ordered} state for $S=1/2$.

\section{Fluctuations around the classical ground state}

Classical (thermal) fluctuations were studied in Refs.\
\onlinecite{peter,Zhito,cghp} such that we make only a few comments
valid for non-zero magnetization and arbitrary anisotropy $\Delta$.
As was shown explicitly for $\Delta = 1$ in Ref.\ \onlinecite{Zhito}, thermal
fluctuations select collinear `up-up-down' (UUD)
configurations at $\langle M \rangle=1/3$ against the other
non-collinear configurations that also minimize the classical
energy, but all UUD configurations have the same spectra of harmonic
fluctuations. Indeed, a local change of variables shows that
the covering-dependent Hamiltonians of classical Gaussian fluctuations
\cite{cghp} are equivalent.

%
%

The role of quantum fluctuations is however radically different.
Now quantum commutation relations have to be preserved and the change
of variables used for the classical case is no longer possible.
To analyze this in more detail, we compute the zero-point
contribution to the ground state energy at $\langle M \rangle = 1/3$
for two different coverings with a $q=0$ and a $\sqrt{3}\times \sqrt{3}$
structure, respectively (the latter can be found e.g.\ in Fig.~1 of
Ref.\ \onlinecite{Zhito}). By writing the spin operators on each
site in terms of bosonic creation and annihilation operators:
\beq
\vec s_l =S \left(\frac{1}{\sqrt{2S}}(a^\dagger_l+a_l), \
i \frac{1}{\sqrt{2S}}(a^\dagger_l-a_l),\
1-\frac{a^\dagger_la_l}{S}\right) \
, \label{hp}
\eeq
we obtain the Hamiltonian:
\beq
H = H_0 + \frac{S}{2} ( H_2 + O( 1/\sqrt{S})) \, ,
\eeq
where $H_2$ is quadratic in creation and annihilation
operators and the $O(1/\sqrt{S})$ part contains higher orders. By
Fourier transforming, we obtain:
\beq
H_2= \frac{J}{2} \sum_{\vec{k}}
      \left( a_{-\vec{k}}^{\dagger i} \, , \, a_{\vec{k}}^i \right)\cdot
      \left( \begin{array}{cc}
  \tilde{M}^+ &  \tilde{M}^- \\
  \tilde{M}^- &  \tilde{M}^+
\end{array}
\right)_{ij} \cdot \left(
\begin{array}{c}
a_{-\vec{k}}^j   \\
a_{\vec{k}}^{\dagger j}
\end{array}
\right) \ , \label{hqtransf}
\eeq
where $\tilde{M}^\pm$ are $3 \times 3$ and $9 \times 9$ matrices
for the $q=0$ and the $\sqrt{3}\times \sqrt{3}$ states since these
coverings have $3$ and $9$ sublattices, respectively. No further
change of variables is possible here since the commutation
relations of the $su(2)$ algebra of the spins have to be preserved.
At $\Delta = 1$ one finds for the zero-point fluctuations
$\frac{1}{2} \sum_{\vec{k}} \omega_{\vec{k}} = JS/3$ and $\approx 0.5643 \, JS$
for the $q=0$ and the $\sqrt{3}\times \sqrt{3}$ state, respectively,
demonstrating the inequivalence of the different coverings at the
quantum level.

\section{Effective model for the Ising limit}

Let us now return to the extreme quantum case $S=1/2$ and
study the anisotropic $XXZ$ limit. For $\Delta \gg 1$, we can
adapt the analysis of Refs.\ \onlinecite{MSC2000,MS2001} of the
Ising model in a transverse field to the $XXZ$ model by replacing
the expansion in the transverse field with an expansion in powers of
$1/\Delta$.

In the Ising limit $\Delta = \infty$, the ground states are those
states where around each triangle two spins point up and one down.
This ground-state space of the Ising model can then be taken as
configuration space for a perturbative treatment of
the $XY$-part of the $XXZ$ Hamiltonian,

For sufficiently large lattices, the lowest non-trivial order is third order,
flipping simultaneously pairwise antiparallel spins around a
hexagon. This is described by an effective Hamiltonian \cite{MSC2000,MS2001}
\bea
H_{\rm eff.} = \lambda \sum_{{\rm hexagon}\ i} \left(
s^+_{i,1} s^-_{i,2} s^+_{i,3} s^-_{i,4} s^+_{i,5} s^-_{i,6} \right. \nn \\
+ \left. s^-_{i,1} s^+_{i,2} s^-_{i,3} s^+_{i,4} s^-_{i,5} s^+_{i,6}
\right) \, , \label{Heff}
\eea
where the spin operator $s^\alpha_{i,j}$ operates at the $j$th site
around hexagon $i$ and $\lambda = 3 J /( 2 \Delta^2)$.
Note that in the Ising-basis the effective Hamiltonian (\ref{Heff}) has
only off-diagonal matrix elements of size $\lambda$.

The configurations of the Ising model can be mapped to dimer
coverings of the dual lattice which in the case of the kagom\'e
lattice is the hexagonal lattice. Now one can use known results for
dimer coverings \cite{Wannier,Wu} to write down
the asymptotic growth law for the number of Ising
configurations ${\cal N}_{\rm conf.}$ on an $N$-site kagom\'e lattice:
\begin{equation}
{\cal N}_{\rm conf.} \propto (1.11372781\ldots)^N \, . \label{Nconf}
\end{equation}

Exploiting results for the related quantum dimer model on the
hexagonal lattice,\cite{MSC2001} Moessner and Sondhi concluded
\cite{MS2001} that the ground state of the effective Hamiltonian
(\ref{Heff}) is of the valence-bond crystal type. To be more precise, the
case studied in Refs.\ \onlinecite{MS2001,MSC2001} corresponds to $\lambda < 0$
whereas we have $\lambda > 0$, but there exist unitary transformations
which change the sign of $\lambda$.\cite{Mpriv} Hence the spectra of the
effective Hamiltonian (\ref{Heff}) are invariant under $\lambda \to -\lambda$.
The three-fold degenerate ground-state wave functions
are sketched in the inset of Fig.\ \ref{figKagM1o3}. Circles in one
third of the hexagons denote resonances between the
two different N\'eel states on the surrounding hexagon; a background
of the remaining third of all spins points in the direction of the field.
Note that these wave functions were argued in \cite{MS2001,MSC2001} to
yield a qualitatively correct description, but they should not be used
for a quantitative analysis. Furthermore, we emphasize
that due to the resonances, these wave functions are of a purely quantum
nature and have no counterparts as unique states of the classical
Heisenberg model.

\begin{figure}
\centerline{\psfig{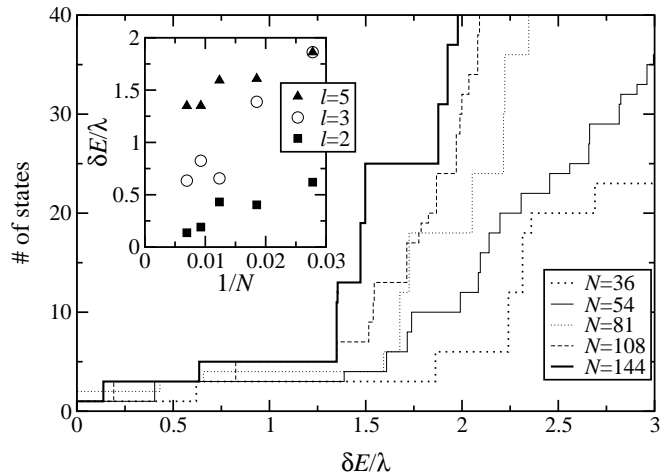}}
\caption{ \label{fig2}
Main panel: Spectra of the effective Hamiltonian (\ref{Heff}) for
$\Delta \to \infty$ with $N=36$, $54$, $81$, $108$ and $144$.
Inset: Scaling of the energy of the $l$th excited state with inverse
system size $1/N$ for some selected levels.
 }
\end{figure}

According to the above, at large $\Delta$ the $\langle M \rangle =
1/3$ state of the $XXZ$ model on the kagom\'e lattice should be
{\it three}-fold degenerate with a {\it gap} to the next
non-magnetic excitations. To check this conclusion and compare it
to Fig.\ \ref{figLL36}, let us look at the spectrum of the
effective Hamiltonian (\ref{Heff}). This effective model has a
substantially reduced Hilbert space (e.g.\ for $N=36$ there are
only 120 states). We can therefore go to larger lattice sizes than
in the full model. Results for kagom\'e lattices with up to
$N=144$ sites are shown in Fig.\ \ref{fig2}.
Additional short cycles wrap around the boundaries
of the lattice for $N\le 27$ and lead to non-generic ground states
of $H_{\rm eff.}$.
Accordingly, systems with $N < 36$ should not be considered and
are not included in Fig.\ \ref{fig2}.

Two features are apparent in Fig.\ \ref{fig2} at
least for the two biggest system sizes ($N=108$ and $144$). Firstly,
there are two further levels above the ground state. The finite-size
dependence of the second excited state is shown by $l=2$ in the
inset of Fig.\ \ref{fig2} and indicates that it is
converging to $\delta E \to 0$ which is consistent with the expected
three-fold degeneracy of the ground state in the thermodynamic limit.
Secondly, there is a huge density of states
emerging for $\delta E \ge 1.3 \lambda$. The finite-size behavior of
the $l=5$ level in the inset of Fig.\ \ref{fig2} indicates that
a gap of the order $\sim 1.2 \lambda$ to these higher excited states
persists in the thermodynamic limit. For $N=108$ and $144$ there
are two further levels in between. It is difficult to extrapolate their
energies to $N \to \infty$, but the behavior of the
$l=3$ excited level in the inset of Fig.\ \ref{fig2} at the largest
values of $N$ suggests that they retain a finite gap in the thermodynamic
limit.
Inspection of the
wave-functions indicates that these additional low-lying levels may arise
from the three classical $\sqrt{3} \times \sqrt{3}$ configurations.

\section{From the Ising limit to the Heisenberg model}

Although the effective Hamiltonian leads to higher degeneracies of
some excited states,
the $N=36$ curves in Figs.\ \ref{figLL36} and \ref{fig2} have a very
similar shape which can be taken as a first indication that the
same scenario as for $\Delta \gg 1$ also applies to $\Delta = 1$.
Comparison of the overall scales leads to an
estimate for the gap in the $\langle M \rangle = 1/3$
sector at $\Delta = 1$ of about $0.04 \, J$.
Furthermore, the total number of Ising configurations is
very close to the number of non-magnetic excitations below the
magnetic gap for $\Delta = 1$ at a given system size
(see also Ref.\ \onlinecite{SCESproc}).
Hence, the growth law (\ref{Nconf}) yields a good approximation also
to the number of non-magnetic excitations in the Heisenberg model ($\Delta=1$).

%
%

\begin{figure}
\centerline{\psfig{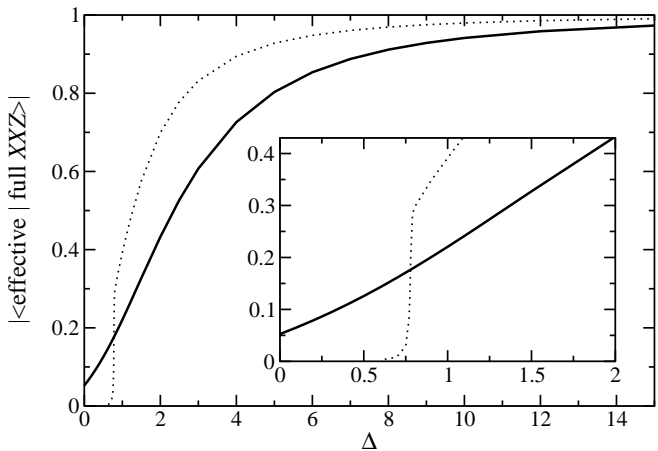}} \caption{
Overlap between the ground state of the full $XXZ$ model and the
effective Hamiltonian with $N=36$ for different values of the anisotropy
parameter $\Delta$ (full line). For comparison we include
the corresponding result for a different model, namely
the $N=36$ triangular lattice \cite{HSR03}
(dotted line) which exhibits a sharp drop around $\Delta = 0.77$.
 \label{fig6} }
\end{figure}

It is instructive to compute the overlap of the wave
function of the full $XXZ$ model, $\vert \hbox{full }XXZ\rangle$,
with the ground state wave function, $\vert
\hbox{effective}\rangle$, of the effective Hamiltonian with the
same number of spins $N$. The analysis of the effective
Hamiltonian implies that one should study only sizes which are
multiples of 9 and that $N=36$ is the smallest which is
representative of the general case. However, $N=36$ is the
biggest system where we have been able to study the full $XXZ$ model and
hence the only case we can discuss. Results for the overlap
$\abs{\langle \hbox{effective} \vert \hbox{full }XXZ\rangle}$ are
shown by the full line in Fig.\ \ref{fig6}. We observe that this
overlap tends to 1 for large values of $\Delta$, as expected.
Furthermore, the overlap remains appreciable even close to the
Heisenberg model ($\abs{\langle \hbox{effective} \vert \hbox{full
}XXZ\rangle} \approx 0.22$ for $\Delta = 1$), in particular if one
considers that the dimension of the symmetry subspace under
consideration is of the order $10^7$. Note further that an
analogous computation of the overlap of the ground state of the
Ising and the full $XXZ$ model at $\langle M \rangle = 1/3$ on the
triangular lattice leads to a sharp drop at $\Delta \approx 0.76$
for a fixed $N$ (see dotted curve in Fig.\ \ref{fig6}
for $N=36$), signaling an instability of the plateau state.\cite{HSR03}
No such sharp drop is observed on the kagom\'e
lattice (full line in Fig.\ \ref{fig6}) which we take as a sign of
absence of phase transitions between $\Delta = \infty$ and
$\approx 0$ in the $XXZ$ model on the kagom\'e lattice at $\langle
M \rangle = 1/3$. In particular, $\Delta = 1$ and $\infty$ should
belong to the same phase.

\section{Discussion and conclusions}

To conclude, we have analyzed the low-energy
spectrum of the kagom\'e $XXZ$ $S=1/2$ model at magnetization
$\langle M \rangle=1/3$. While the existence of a magnetization plateau
is clear, the nature of the non-magnetic excitations over the ground state
is more difficult to clarify. We have argued by different
techniques that the ground state has an order of the valence-bond crystal type,
{\it i.e.}\ the ground state is three-fold degenerate and there is a small gap
to all higher excitations. While in the case $\Delta \gg 1$ this scenario
is derived from a mapping to an effective Hamiltonian,%
\cite{MSC2000,MS2001,MSC2001} our numerical data indicates that it persists
down to the isotropic limit $\Delta = 1$.

One of the key differences between the present case and $\langle M
\rangle=0$ lies in the unrenormalized classical thermal and
quantum fluctuations. In the absence of a magnetic field, they are
equivalent regarding the lifting of degeneracy of configurations
with soft modes (planar configurations). However, for $\langle M
\rangle=1/3$, classical thermal fluctuations select the collinear
UUD configurations and the weight in the free energy of any UUD
covering is equivalent at the harmonic level. Because of
commutation relations that have to be preserved at the quantum
level, the zero-point corrections over the UUD configurations are
not any more equal. This is the first indication that a spin liquid phase is
less likely to appear than for the $\langle M \rangle=0$ case.

For $S=1/2$ and $\Delta = 1$ we find, for $\langle M \rangle = 1/3$, an
exponential number of non-magnetic excitations  below the magnetic
gap which are reminiscent of the classical degeneracy; just as for
$\langle M \rangle = 0$.\cite{lhuillier03,waldtmann98,Mila} In
the latter case the macroscopic number of non-magnetic excitations
has been taken as evidence for a completely disordered ground
state. Here, however, we find evidence for a further small gap,
separating the continuum of states from a ground state, which has
long-range order of valence-bond crystal type. We remark that
the $N=36$ spectrum \cite{waldtmann98} suggests that candidates for
valence-bond ordered states for $\langle M \rangle = 0$ would have
a larger unit cell than the state above. Hence, we believe that
the issue of order at very low energies in the $S=1/2$ KHAFM 
at $\langle M \rangle = 0$ remains a challenging problem.

\begin{acknowledgments}

We would like to thank the Rechenzentrum,
TU Braunschweig for allocation of CPU time on the compute-server {\tt cfgauss}
and J.\ Sch\"ule for technical support. We are grateful to H.-U.\
Everts, C.\ Lhuillier, F.\ Mila and R.\ Moessner for useful
discussions. This work was financially supported by ECOS-Sud
and Procope-Egide exchange programs as well as by the
Deutsche Forschungsgemeinschaft (project Ri615/12-1). The research
of D.C.C.\ and M.D.G.\ was partially supported by CONICET and
Fundaci\'on Antorchas, Argentina (grant No.\ A-13622/1-106).

\end{acknowledgments}


\end{document}